\documentclass[pra,showpacs,twocolumn]{revtex4}

\usepackage{amsmath}    % need for subequations
\usepackage{graphicx}   % need for figures
\usepackage{verbatim}   % useful for program listings
\usepackage{color}      % use if color is used in text
\usepackage{subfigure}  % use for side-by-side figures
\usepackage{hyperref}   % use for hypertext links, including those to external documents and URLs

\begin{document}

\title{Theory of ultrafast autoionization dynamics of Fano resonances}
\author{W.-C. Chu and C. D. Lin}
\date{\today}
\affiliation{J. R. Macdonald Laboratory, Department of Physics, Kansas State University, Manhattan, Kansas 66506, USA}

\begin{abstract}
We study atomic autoionization processes in the time domain. With the emerging attosecond extreme vacuum ultraviolet and soft x-ray
pulses, we first address how to characterize the time evolution of the decay of a discrete state into a degenerate continuum. A short
pump beam generates a number of resonance states in a series and the nearby background continuum, and the resultant wave packet
evolves with time until the full decay of the bound states. Taking the $2pns (^1P^o)$ resonance series embedded in the
$2s \epsilon p (^1P^o)$ continuum in beryllium atom as an example, the time evolution of the autoionizing wave packet in energy
domain and in coordinate space is calculated
and analyzed, where Fano profiles build up in the photoelectron energy during the process. A proposed pump-probe scheme assumes that
the probe beam ionizes the $2s$ inner electron in the wave packet. The lifetimes of the resonances and the photoelectron energy
distribution can be obtained from the ionization yield versus the time delay of the probe.
\end{abstract}

\maketitle

% INTRODUCTION %%%%%%%%%%%%%%%%%%%%%%%%%%%%%%%%%%%%%%%%%%%%%%%%%%%
\section{Introduction}
\label{sec:introduction}

Scattering processes -- photoionization, electron collision, etc. -- have been important tools for studying physical
systems since the early days of the development of atomic theories. Autoionization, first observed by Beutler~\cite{beutler} in
the photoabsorption of rare gas atoms, is a phenomenon where the energy spectra is characterized by asymmetric resonance peaks, which
were explained by Fano in his seminal paper in 1961~\cite{fano}. In this paper, asymmetric resonance in photoabsorption occurs
in the energy region where a discrete state is embedded in the continuum. Photoabsorption populates both the discrete state and the
continuum states. The interaction of the discrete state with the continuum results in the "autoionization" of the former, where the
discrete state interferes with the directly populated continuum states to form asymmetric Beutler-Fano profile. Today Fano
resonances have been used to describe a wealth of physical phenomena in atomic, molecular and optical systems, condensed matter
systems\cite{wick14,wick15}, nanostructures\cite{mirosh} as well as many other fields. Essential to Fano's theory is that
each resonance can be characterized by three parameters, the "unshifted" resonance position $E_r$, the shape parameter $q$ and the
resonance width $\Gamma$. The resonance width characterizes the decay time or the lifetime of the discrete state. Since the typical
resonance width of an atomic or molecular system is in the order of 10$^{-1}$~eV, which corresponds to the lifetime of as short as a
few femtoseconds, almost all measurements on these resonances have been carried out in the energy domain, using high-resolution
spectroscopy. Today a large set of experimental and theoretical data of these resonances have been collected, and they provide
important structural information of the systems.

In the past decade, with the advent of laser technology, single attosecond pulses (SAP) are becoming available\cite{krausz}. Such
pulses opened up a new research platform for studying electronic dynamics, such as the autoionization dynamics. Indeed, the first SAP
experiment in atomic system was to determine the lifetime of the Auger decay of an $M$-shell hole in Kr created by extreme vacuum
ultraviolet (XUV) attosecond pulse by Drescher \textit{et al}~\cite{drescher}. For Auger decay, the resonance profile is mostly
symmetric, and thus only the decay width or the lifetime of the inner-shell hole was determined. In Ref.~\cite{drescher}, the
lifetime is determined by creating the inner-shell hole in the presence of an IR laser. By changing the time delay between the XUV
pulse with the IR, photoelectron spectra can be used to "deduce" the lifetime of the inner-shell hole -- based on the approximate
"streaking" theory\cite{kitzler}. Such streaking theory has been generalized to Fano resonances by Zhao and Lin~\cite{zhao} as well
as by others~\cite{wickenhauser,smirnova}. However, the photoelectron spectra from such XUV+IR setup are very complicated and only
the lifetime of the Fano resonance can be extracted.

Due to the insufficient fluence of the SAP's so far, existing measurements using attosecond pulses for Auger decay or autoionization have
all been carried out using the XUV+IR pump-probe setup where the two pulses actually overlap in time\cite{drescher,gilbertson,wang}.
Rigorously speaking, the IR is not probing the resonances created by the attosecond XUV pulse in these experiments since the
photoelectron is created in the IR field. With improved technology in attosecond pulse generation and the recent advances in
X-ray free-electron lasers (XFEL), we envision that intense attosecond pulses will become available in the near future, such that a
true time-dependent investigation of the autoionization process will be possible. In this article, we analyze
the time evolution a Fano resonance, or the autoionization dynamics of the system in the field-free condition before the discrete
state fully decays into the continuum. For example, if the discrete state has a lifetime of 20 fs while the pump pulse is only 1.5
fs, we ask what is the time evolution of the wave packet associated with the decay process after the pump pulse, and how the energy
profile eventually evolves into what is characterized by the Fano resonance shape.

We comment that the time evolution of a Fano resonance has been studied in several calculations by solving the time-dependent
Schr\"odinger Equation of helium or a model atom~\cite{mercouris,lindroth,wickenhauser,cavalieri}, with or without the IR probe. These
studies did not address how to retrieve the time evolution of the resonance profiles. Similarly, autoionization of double
Rydberg states in the time scale of picoseconds has been studied\cite{jones}. In the meanwhile, the interaction of a doubly excited
state with a whole Rydberg series also falls into the similar category\cite{gallagher1,gallagher2}. Understanding the time evolution
of the two-electron wave packets and the issue of how to probe their evolution lie at the heart of understanding and controlling
two-electron dynamics.

In this article, we use two pulses that are separate in time. The probe is always placed strictly after the pump so that the
field-free evolution between the pulses can be defined. The physical process is understood roughly as the decaying of the bound state
to the continuum due to the interaction between them. In Sec.~\ref{sec:theory}, starting with Fano theory which is for stationary
states, we construct the time-dependent wave function of the system which describes the time evolution of the electron wave packet
after the pump pulse. The wave function is recasted in the basis set of the bound state and the continuum states, where it
contains the time information for the bound and the continuum components. The wave packet for the continuum part is shown to
evolve to the typical Fano resonance profile at long time. We thus define the continuum wave packet at short time as the
time-dependent Fano profile of the autoionization. Since the pump pulse is short, it inevitably spans a broad-band in the energy
spectra. Assuming that the background energy spectrum generated by the pump is described by a Gaussian distribution, we generalize the
theory to include all the resonances within the bandwidth. In Sec.~\ref{sec:application} we apply our theory to the resonances
in beryllium to investigate wave packet dynamics after the pump pulse. In Sec.~\ref{sec:probe} we demonstrate how to probe the
time-dependent Fano resonances, using the example in beryllium given in Sec.~\ref{sec:application}. A short summary and future
outlook is given in Sec.~\ref{sec:conclusions}.

% THEORY %%%%%%%%%%%%%%%%%%%%%%%%%%%%%%%%%%%%%%%%%%%%%%%%%%%
\section{Theory}
\label{sec:theory}

% A. Fano's theory
\subsection{Fano's theory of resonance} \label{sub:fano}

In a multi-electron atomic system, the high-lying discrete states are embedded in the continuous spectrum if these discrete energies
are higher than the binding energy. Both the bound and continuum states are described in terms of configurations. According to
Ref.~\cite{fano}, with one bound state and its nearby continuum, the system is governed by the total Hamiltonian given by
\begin{align}
\langle \alpha |H| \alpha \rangle &= E_r, \label{eq:h_bound} \\
\langle \beta_E |H| \alpha \rangle &= V_E, \label{eq:h_mix} \\
\langle \beta_{E'} |H| \beta_E \rangle &= E \delta (E'-E), \label{eq:h_cont}
\end{align}
where $|\alpha \rangle$ and $| \beta_E \rangle$ are bound and continuum configurations respectively. The off-diagonal
terms $V_E$ represent the mixing strength between the bound and the continuum. The diagonal terms $E_r$ and $E$ are bound
and continuum energies respectively.

The eigenstates $| \psi_E \rangle$ of the atom are obtained by diagonalizing the Hamiltonian in Eq.~\ref{eq:h_bound}-\ref{eq:h_cont}
and are given by Fano as
\begin{equation}
| \psi_E \rangle = a_E | \alpha \rangle + \int{ b_{EE'} | \beta_{E'} \rangle \text{d}E' }, \label{eq:eigen-config}
\end{equation}
where the coefficients are
\begin{align}
a_E &= \frac{\sin{\Theta}}{\pi V}, \label{eq:def_a} \\
b_{EE'} &= \frac{1}{\pi} \frac{\sin{\Theta}}{E-E'} - \delta (E-E') \cos{\Theta}, \label{eq:def_b} \\
\Theta &\equiv -\arctan{\frac{\pi V^2}{E-E_r}}. \label{eq:def_Theta}
\end{align}
Here we assume that $V_E=V$ is a real constant in the narrow energy region for each Fano resonance, provided that the bound and
the continuum states are in real representation.

Originally Fano's theory was applied to photoabsorption processes. The transition amplitude is $c_E \equiv \langle \psi_E |T| g
\rangle$ where $| g \rangle$ is the ground state, $| \psi_E \rangle$ is the excited state of energy $E$, and $T$ is the photon
transition operator. The states $| g \rangle$ and $| \psi_E \rangle$ are eigenstates of the atomic Hamiltonian $H$, so the
coefficients $c_E$ do not change with time. In Ref.~\cite{fano}, the continuum wave functions are taken to be real standing
waves. The absorption cross section is proportional to $|c_E|^2$, which contains the features of Fano profiles.
Until recently, all photoabsorption measurements are performed using light sources of hundreds of picoseconds or longer, where they
can be considered as monochromatic. Thus, $| \psi_E \rangle$ of a single $E$ is promoted in each measurement. To trace
out a Fano resonance, monochromatic lights are tuned across the resonance while the cross section for each energy point is recorded.

Although Fano's theory was applied to photoabsorption originally, it can be trivially generalized to study photoelectron angular
distributions. While the continuum functions in Eq.~\ref{eq:eigen-config} are taken to be real standing waves, $| \psi_E \rangle$
is an eigenstate of the Hamiltonian. To obtain the scattering amplitude for an electron with momentum $\vec{k}$, one only needs to
project the standing wave onto the momentum eigenstate $\vec{k}$ of the photoelectrons, in energy-normalized form given by
\begin{equation}
\psi_{\vec{k}} (\vec{r}) = \sqrt{\frac{2}{\pi k}} \frac{1}{r} \sum_{lm} {i^l e^{i \eta_l} u_l (kr) Y_{lm} (\hat{r}) Y_{lm}^* (\hat{k}) }.
\end{equation}
Here $u_l(kr)$ are the radial wave function of the electron taken as real standing waves as in Fano's theory, $\eta_l$ are the total
scattering phase shifts, and the $Y$'s are spherical harmonic functions.

Following the notation in Ref.~\cite{fano}, the width of the resonance is defined by $\Gamma \equiv 2 \pi V^2$. Assuming that
$\langle \beta_E |T| g \rangle$ is energy-independent near the resonance energy $E_r$, the $q$-parameter is defined by
\begin{equation}
q \equiv \frac {\langle \alpha |T| g \rangle} {\pi V \langle \beta_E |T| g \rangle}. \label{eq:q_fano}
\end{equation}
Without the flat-background and the constant $V$ assumptions, $| \alpha \rangle$ will be replaced by the "modified bound state"
defined in Ref.~\cite{fano} which contains a term based on uneven $V_E$ and $| \beta_E \rangle$ near $E_r$. We assume this modification
is negligible in general cases and consider the $q$-parameter as defined in Eq.~\ref{eq:q_fano}. Fano's theory predicts the resonance
profile in a simple form, where the associated parameters $E_r$, $\Gamma$, and $q$ can be extracted by fitting the form to the measured
profile.

With the advent of available light pulses of a few femtoseconds or shorter, one can use such a short pulse to ionize the atom. After
the pump pulse is over, the time-dependent wave function of the atom is given by
\begin{equation}
| \Psi(t) \rangle = c_g e^{-iE_gt} | g \rangle + \int{ c_E e^{-iEt} | \psi_E \rangle \text{d}E } \label{eq:Psi_eigen}
\end{equation}
where in a single pulse, the integral in Eq.~\ref{eq:Psi_eigen} covers a bandwidth more than the width of a resonance. Such a
resonance can be measured, long after the pump, in a  high-resolution spectrometer. This amounts to projecting out the time-dependent
wave function $| \Psi(t) \rangle$ over a long time. Thus the whole Fano resonance is generated in a single pulse and its profile is
given by $|c_E|^2$.

We now consider the situation where the pump pulse is shorter than the lifetime of a Fano resonance. One would like to "see" the
dynamics of the autoionization. However, such information is not contained in $|c_E|^2$ since $| \psi_E \rangle$ are eigenstates of
the atomic Hamiltonian and the coefficients $c_E$ are constant when the pump field is absent. Thus in Eq.~\ref{eq:Psi_eigen}, the time
information is hidden and not recognized, even though we "know" that autoionization is happening!

% Theory B. Time-dep
\subsection{Time-dependent wave function in configuration basis} \label{sub:time-dep}

It is convenient to define the excited wave function $|\Psi_{ex}(t)\rangle$ as the total wave function $|\Psi(t)\rangle$ excluding
the ground state part since the ground state does not participate in autoionization. Instead of using eigenstates as expressed in
Eq.~\ref{eq:Psi_eigen}, we alternatively expand the excited wave function in configuration basis as
\begin{equation}
|\Psi_{ex}(t)\rangle = d_{\alpha}(t) |\alpha\rangle + \int{ d_E(t) |\beta_E\rangle \text{d}E }. \label{eq:Psi_single}
\end{equation}
Since $|\alpha\rangle$ and $|\beta_E\rangle$ are not eigenstates of the atomic Hamiltonian, the coefficients $d_{\alpha}(t)$ and $d_E(t)$
are time-dependent. One expects that at long time (compared to the lifetime), the coefficient $d_{\alpha}(t)$ will go to zero and all
the information of the total wave function will be contained in $d_E(t)$. However, the time evolution of these coefficients cannot
be measured in laboratory -- they change in the time scale of femtoseconds or less, while laboratory electronics can only measure
changes of the order of picoseconds or longer. Here we define $|d_E(t)|^2$ as the "time-dependent Fano resonance profile", or in
short, "resonance profile". Below we discuss how these coefficients behave. The measurement issues will be addressed later.

Assuming that the initial values of $d_{\alpha}(t)$ and $d_E(t)$ are known, with the aid of Eq.~\ref{eq:eigen-config}-\ref{eq:def_Theta},
the time-dependent coefficients are solved and given by
\begin{align}
d_{\alpha}(t) &= \left[ d_{\alpha}^{(0)} e^{-\frac{\Gamma}{2} t} + \int{ d_E^{(0)} g_E(t) \text{d}E } \right] e^{-iE_rt},
\label{eq:exact_bound} \\
d_E(t) &= \left[ d_{\alpha}^{(0)} g_E(t) + \int{ d_{E'}^{(0)} f_{EE'}(t) \text{d}E' } \right] e^{-iE_rt}\notag \\
&+ d_E^{(0)} e^{-iEt}, \label{eq:exact_cont}
\end{align}
where the functions $g_E(t)$ and $f_{EE'}(t)$ are defined by
\begin{align}
g_E(t) &\equiv \frac{V}{E- E_r+i\Gamma/2} \left[ e^{-i (E-E_r) t} - e^{-\frac{\Gamma}{2} t} \right], \label{eq:exact_g}\\
f_{EE'}(t) &\equiv \frac{V}{E-E'} \left[ g_E(t) - g_{E'}(t) \right]. \label{eq:exact_f}
\end{align}
The superscript $(0)$ denotes the initial values, i.e.,
\begin{align}
d_{\alpha}^{(0)} &= \langle \alpha | \Psi(0) \rangle , \\
d_E^{(0)} &= \langle \beta_E | \Psi(0) \rangle .
\end{align}
The solutions in Eq.~\ref{eq:exact_bound}-\ref{eq:exact_f} are calculated exactly for a given set of parameters $d_{\alpha}^{(0)}$,
$d_E^{(0)}$, $E_r$, and $\Gamma$, where these parameters can be either obtained by running a separate program such as Time-dependent
Schr\"{o}dinger Equation (TDSE) code or time-dependent perturbation calculation, based on the prior knowledge of the pump, or
extracted from the experimental values in the time-integrated spectrum. In the latter case, with $q$ defined by Eq.~\ref{eq:q_fano}
and rewritten as
\begin{equation}
q \equiv \left. \frac{d_{\alpha}^{\left( 0 \right)}}{\pi Vd_E^{\left( 0 \right)}} \right|_{E=E_r}, \label{eq:q_conf}
\end{equation}
the initial coefficient $d_{\alpha}^{(0)}$ can be extracted since the value of $q$ and the background magnitude of $d^{(0)}_E$ can
be obtained from the Fano profile after long time (the phase of $d^{(0)}_E$ can be deduced if photoelectron angular distributions
are measured). Equations~\ref{eq:exact_bound}-\ref{eq:exact_f} are the backbone of our model.

Note that the initial coefficients $d_{\alpha}^{(0)}$ and $d_E^{(0)}$ are in general complex values while $q$ is real. Within the
narrow range of a resonance width, the energy $E$ in $d_E^{(0)}$ is close to the bound state energy $E_r$ such that the phases of
$d_{\alpha}^{(0)}$ and $d_E^{(0)}$ are assumed to be the same in the neighborhood of the resonance, and the corresponding $q$,
representing the ratio between $d_{\alpha}^{(0)}$ and $d_E^{(0)}$, is real. When $d_E^{(0)}$ spans a wide energy range covering many
resonances, its phase varies smoothly with $E$, but in each resonance region, the statement above is still true, and each $q$ is real.

The decay lifetime of the bound state is defined as $T \equiv 1/\Gamma$. Rigorously speaking, the dynamics of the autoionization is not
only about the decay of the bound state, but also its coupling to the continuum (See Eq.~\ref{eq:Psi_single}). The wave packet can
propagate from the continuum to the bound or the other way around, or between continuum states of different energies through the
bound state. Thus, the lifetime $T$, determined by the interaction matrix $V$, characterizes the time scale of the whole event.

We can generalize Eq.~\ref{eq:exact_bound}-\ref{eq:exact_f} to include multiple resonances by applying Fano's theory for many bound
states~\cite{fano} and assuming that higher-order perturbation by $V$ is negligible. The excited wave function is
\begin{equation}
\Psi_{ex}(t) = \sum_n{d_n(t) | \alpha_n \rangle} + \int{ d_E(t) |\beta_E\rangle \text{d}E }, \label{eq:Psi_multi}
\end{equation}
with the coefficients given by
\begin{align}
d_n(t) &= \left[ d_n^{(0)} e^{-\frac{\Gamma_n}{2} t} + \int{ d_E^{(0)} g_{nE}(t) \text{d}E } \right] e^{-iE_nt},
\label{eq:multi_bound} \\
d_E(t) &= \sum_n { \biggl\{ \left[ d_n(t) g_{nE}(t) + \int{ d_{E'}^{(0)} f_{nEE'}(t) \text{d}E' } \right] } \notag \\
& \times e^{-iE_nt} \biggr\} + d_E^{(0)} e^{-iEt}, \label{eq:multi_cont}
\end{align}
where the label $n$ is the index for the $n$th bound state. The $g$-function and the $f$-function, now
subscripted by $n$, are calculated independently for each resonance with the associated parameters as given by
Eq.~\ref{eq:exact_g}-\ref{eq:exact_f}.

To simulate the effect of the finite range of $V$ while keeping the calculation simple, the profiles can be further adjusted by
multiplying each "resonance term" in Eq.~\ref{eq:multi_cont} (in the $\{\}$ bracket) by a Gaussian window function to limit the range
of each resonance. For an isolated resonance, the effect of the window function is minimum or unnecessary because the window function
mainly removes "the wings" but not the center of the resonance, where the central part is the significant and informative part. On the
other hand, when dealing with many resonances, without the window functions, the wings may be overly extended and unnaturally perturb
the nearby resonances. In the latter case, multiplying the window functions is an effective and efficient solution.

%%%%%
\subsubsection{Flat initial continuum (IC) distributions}

To study the qualitative behavior of the wave function, we first assume that the initial profile is flat, i.e. $d_E^{(0)} = const$.
This initial condition, dubbed IC$_1$, is valid near the resonance positions if the pump bandwidth is broad enough relative to the
resonance widths so that the populated profile is nearly flat around $E_r$. For conceptual presentation, only the
single-resonance formulae Eq.~\ref{eq:exact_bound}-\ref{eq:exact_f} are referred here, while the idea is the same for the
many-resonance case. Defining the scaled time $s$ by $s \equiv \Gamma t$ and scaled energy $\epsilon$ by $\epsilon \equiv 2 (E-E_r)
/ \Gamma$, the bound and the continuum coefficients under IC$_1$ are
\begin{align}
d_{\alpha}(s) &= d_{\alpha}^{(0)} (1-\frac{i}{q}) e^{-\frac{1}{2} s}, \label{eq:IC1_bound}\\
d_{\epsilon}(s) &= d_{\epsilon}^{(0)} \frac{1}{\epsilon + i} \left[ (q+\epsilon) e^{-\frac{i}{2} \epsilon s} - (q-i) e^{-\frac{1}{2} s} \right], \label{eq:IC1_cont}
\end{align}
where in the scaled energy, q-parameter becomes
\begin{equation}
q \equiv \frac {d_{\alpha}^{(0)}} {\pi V d_{\epsilon}^{(0)}}, \label{eq:q_scaled}
\end{equation}
and the factor $e^{-iE_rt}$ disappears because $\epsilon$ is measured relative to the resonance energy. While the coefficients given by
Eq.~\ref{eq:IC1_bound}-\ref{eq:IC1_cont} represent the evolution for $s>0$, their limits at $s \to 0$ are different from the initial
coefficients $d_{\alpha}^{(0)}$ and $d_{\epsilon}^{(0)}$ that we have defined at $s=0$. This discontinuity means an immediate jump of the
wave function right after the evolution starts. This unphysical behavior is due to the integral of $g$-function in Eq.~\ref{eq:exact_g} and
$f$-function in Eq.~\ref{eq:exact_f} over infinite energy range at
very small times, where the contributions from $E$ much higher or lower than $E_r$ are misleadingly counted in, while in practice,
only the energy range comparable to $\Gamma$ should be considered. Nevertheless, the large-time behavior is reliable since both
$g$- and $f$-functions are compressed more and more into the near-resonance region with increasing time.

\begin{figure}[htb]
\centering
\includegraphics[width=\linewidth]{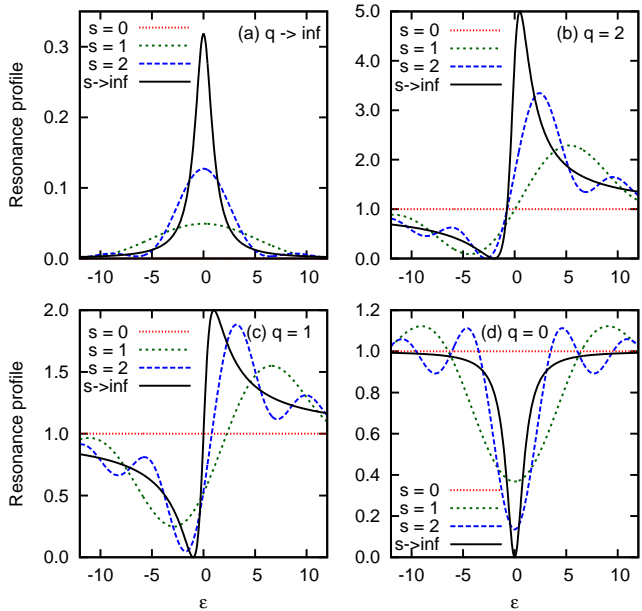}
\caption{The resonance profile $| d_{\epsilon}(s) |^2$ calculated by Eq.~\ref{eq:IC1_cont} for (a) $q \to \infty$, (b) $q=2$, (c)
$q=1$, and (d) $q=0$. For (b)-(d), $| d_{\epsilon}^{(0)} |^2$ is normalized to 1. In all cases, the profile builds up with time with
small oscillations in energy before reaching the final Fano profile.}
\label{fig:IC1}
\end{figure}

In the $s \to \infty$ limit in Eq.~\ref{eq:IC1_bound}-\ref{eq:IC1_cont}, the bound state decays to zero, but the continuum recovers
the general form of Fano profile,
\begin{equation}
\lim_{s \to \infty} |d_{\epsilon}(s)|^2 = |d_{\epsilon}^{(0)}|^2 \frac {| \epsilon+q |^2} {\epsilon^2+1}. \label{eq:IC1_fano}
\end{equation}
We emphasize that Eq.~\ref{eq:IC1_fano} is a mathematical consequence. It cannot be obtained from the measurement since continuum
configurations are not eigenstates of the total atomic Hamiltonian. In the special case where $d_{\alpha}^{(0)} = 1$ and
$d_{\epsilon}^{(0)} = 0$, the system starts with only bound state, and Lorentz profile will be recovered as $s \to \infty$, i.e.,
\begin{equation}
\lim_{s \to \infty} | d_{\epsilon}(s) |^2 = \frac{1}{\pi} \frac{1}{\epsilon^2 +1}. \label{eq:IC1_lorentz}
\end{equation}
Fig.~\ref{fig:IC1} shows the spectra at different times for $q \to \infty$, $q=2$, $q=1$, and $q=0$. Each sub-figure demonstrates how
its spectrum morphs into Fano profile. A feature found in the profile is the oscillation in the wings during the evolution, which
shrinks its energy period when time increases, and disappears as $s \to \infty$. This oscillation, due to the phase between the two
exponential terms in Eq.~\ref{eq:IC1_cont}, is not seen in the conventional time-integrated spectra. Although analytically simple,
the IC$_1$ simplification is conceptually rich and offer an easy visualization for the autoionization process.

%%%%%
\subsubsection{Gaussian initial continuum distributions}
\label{sub:IC2}

\begin{figure}[htb]
\centering
\includegraphics[width=\linewidth]{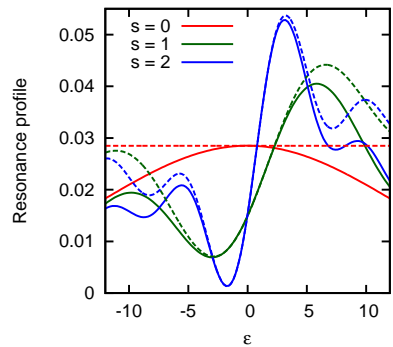}
\caption{The resonance profile for IC$_1$ and IC$_2$ initial conditions. The solid curves are carried out with the IC$_2$ initial
condition defined by $q=1$ and the FWHM bandwidth of $15\Gamma$, or $D=9\Gamma$. The dashed curves are carried out under IC$_1$.}
\label{fig:IC2}
\end{figure}

The condition IC$_1$ is valid only in the vicinity of a resonance and not applicable to the whole energy range. When a short pump
pulse is used to initiate the system, one should cover its whole bandwidth and the state vectors must be properly normalized. Here we
consider the normalization where the total probability for the bound component $\sum_n { |d_n(t)|^2 }$ plus the continuum component
$\int{ |d_E(t)|^2 \text{d}E }$ is set to 1.0, where the ground state $| g \rangle$ is not considered since it is not involved in the
autoionization. This normalization is automatically fulfilled by the exact solution once the initial coefficients are normalized.
Additionally, we assume that the pump pulse has a Gaussian envelope, resulting in a Gaussian profile in energy. The initial
coefficients are given by
\begin{align}
d_n^{(0)} &= \gamma q_n \pi V_n \exp{ \left[ -\frac{1}{2} \left( \frac{E_n-\omega}{D} \right) ^2 \right] }, \label{eq:IC2_bound} \\
d_E^{(0)} &= \gamma \exp{ \left[ -\frac{1}{2} \left( \frac{E-\omega}{D} \right) ^2 \right] }, \label{eq:IC2_cont}
\end{align}
where
\begin{equation}
\frac{1}{\gamma ^2} \equiv \sqrt{\pi} D + \sum_n { (q_n\pi V_n)^2 \exp{ \left[ -\left( \frac{E_n-\omega}{D} \right) ^2 \right] }}.
\label{eq:IC2_gamma}
\end{equation}
This initial condition is dubbed IC$_2$. The initial continuum state has a Gaussian energy distribution centered at $\omega$ with
standard deviation $D$ ($\text{FWHM}=1.665D$). The initial coefficients are determined by the pump width and the resonance
parameters $E_n$, $\Gamma_n$ (or equivalently, $V_n$), and $q_n$. For $D \gg \Gamma_n$, IC$_2$ is reduced back to IC$_1$ near the
$n$th resonance. The comparison between the profiles under the IC$_1$ and IC$_2$ is plotted in Fig.~\ref{fig:IC2} for $q=1$, where the
bandwidth (FWHM) for IC$_2$ is 15$\Gamma$. As seen in Fig.~\ref{fig:IC2}, with the modification from IC$_1$ to IC$_2$, the $s=0$
profile changes from a flat background to a Gaussian shape, and the $s=1$ and $s=2$ profiles change similarly. However, the featuring
structures are preserved. The modification should be slight if the bandwidth is much wider than $\Gamma$, which is true for a
sufficiently short pump.

%%%%%
\subsubsection{Growth and decay of a bound state}
\label{sub:decay}

\begin{figure}[htb]
\centering
\includegraphics[width=\linewidth]{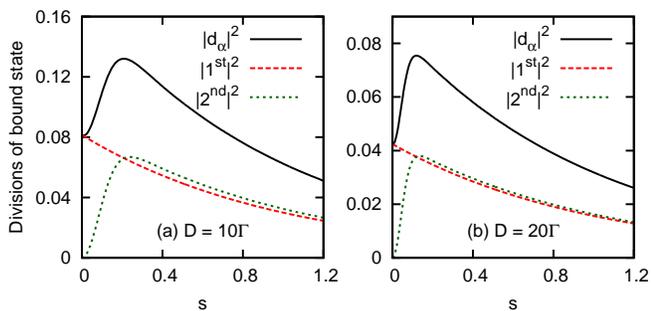}
\caption{The bound state coefficient defined by Eq.~\ref{eq:exact_bound} and its compositions for $q=1$ and (a) $D=10\Gamma$, and
(b) $D=20\Gamma$, where $D$ is the standard deviation of the pump beam (See Sec.\ref{sub:decay}). When $D$ increases
relative to $\Gamma$, the rising portion of the second term at the beginning of time gets shorter, and in the limit
$D \gg \Gamma$, Eq.~\ref{eq:IC1_bound} (IC$_1$) is recovered.}
\label{fig:bound}
\end{figure}

Considering an isolated resonance, in IC$_1$, the total bound state is an exponential decay function of time as described by
Eq.~\ref{eq:IC1_bound}. However, under the normalization condition of IC$_2$, which is more realistic, the bound state decay
is not necessarily monotonic. For the IC$_2$ solution, the bound state coefficient in Eq.~\ref{eq:exact_bound} and its two terms
are shown in Fig.~\ref{fig:bound} in their absolute squares. The two panels are for two different initial bandwidths of the continuum
centered at the resonance. The first term, which comes from the initial bound state, is always an exponential decay. The second term,
which comes from the initial continuum, goes through a rising period up to about $s=0.2$ in Fig.~\ref{fig:bound}(a) and $s=0.1$ in
Fig.~\ref{fig:bound}(b) before it decays. This rising period will shorten if the initial bandwidth increases or the pump duration
shortens. In the limit of infinite bandwidth, both the first and the second terms will be monotonic decay functions, and the solution
in IC$_1$ will be recovered. Note that even in a situation where the length of the rising period is comparable to the lifetime,
such as what Fig.~\ref{fig:bound}(a) shows, the total bound state population will eventually be dominated by the exponential decay.
This growth part of the bound state is usually much shorter than the decay lifetime, so to "watch" this detail structure, a higher
temporal resolution is required.

In the special case for $d_{\alpha}^{(0)}=0$, the pump pulse does not directly populate the bound state of the system. The $q$-parameter
defined by Eq.~\ref{eq:q_conf} is 0. However, this definition is based on the assumption that $V$ and the continuum background are
constant near the resonance, and it excludes the modification term for the bound state. In an actual situation where no original bound state
$|\alpha\rangle$ was populated, the theory needs to be modified to include higher order terms. Nonetheless, even if we neglect the initial
population of the original and the modified bound state, the time-dependent probability as given by Eq.~\ref{eq:exact_bound} will still
have contribution from the initial continuum population $d_E^{(0)}$. In other words, we will still have the "decay" of the bound state even if
it was not populated by the pump pulse. This somewhat awkward result is best understood in the following way. First recall that the
initial bound state population depends on the transition operator connecting it to the ground state, while the "configuration interaction"
$V$ which connects the bound state to the continuum is always present. This "configuration interaction" feeds the bound state from
the continuum, but the continuum states also draw the bound state component and take it to large distances. If these continuum states
were substituted by highly excited bound states, the outer electron would return to recollide with the inner electron(s) and repopulate
the bound state. This would correspond to situations studied in Ref.~\cite{gallagher1,gallagher2,jones}. To initiate the oscillation, the
bound state does not have to be excited by the pump pulse. The ever present "configuration interaction" is the one which is responsible
for the oscillation. Such "interaction" is the property of the Hilbert space but not of the specific excitation mechanism.

% APPLICATION %%%%%%%%%%%%%%%%%%%%%%%%%%%%%%%%%%%%%%%%%%%%%%%%%%%
\section{Application to resonances in Be} \label{sec:application}

In atomic systems, autoionization is understood only in the multiple-electron picture where electron-electron correlation is considered.
Our theory is applied here to calculate the autoionizing wave packet of the $2p4s (^1P^o)$ resonance and of the $2pns (^1P^o)$
($n=3$ to 9) resonance series embedded in the $2sEp (^1P^o)$ continuum in beryllium atom, generated by the photoionization from the
$2s^2 (^1S^e)$ ground state. For a short pump, we assume that the initial continuum distribution can be approximated by a Gaussian
distribution. Only the two outer electrons in beryllium are assumed active, and the $1s^2$ core electrons are frozen. The bound orbitals
are Slater type orbitals where their analytical forms are determined by fitting the numerical Hartree-Fock calculations. The continuum
basis functions are calculated with the model potential given in the literature~\cite{buendia} for $l=1$ standing waves~\cite{joachain}.
The resonance parameters are taken from the experiment of Wehlitz \textit{et al}~\cite{wehlitz}, where $E_r$ is 2.789~eV above the
ionization threshold, $q=-0.52$, and $\Gamma = 0.174$~eV, or $T=3.78$~fs. The pump beam has the carrier frequency right at the resonance
and the duration of 2~fs, or mean bandwidth of 0.912~eV.

For the $2p4s$ resonance case, the excited wave function, Eq.~\ref{eq:Psi_single}, is given by
\begin{align}
\Psi_{ex} (\vec{r_1},\vec{r_2};t) &= d_{2p4s}(t) \phi_{2p4s} (\vec{r_1},\vec{r_2}) \notag\\
&+ \int{ d_{E'}(t) \phi_{2sE'p} (\vec{r_1},\vec{r_2}) \text{d}E' }, \label{eq:Psi_2p4s}
\end{align}
where the $\phi$-functions are constructed by corresponding bound orbitals and continuum waves, and symmetrized
between the two electrons. The one-electron density is defined by
\begin{equation}
\rho(r_1,t) = \iiint{ | \Psi_{ex} (\vec{r_1},\vec{r_2};t) |^2 r_1^2 r_2^2 \text{d}\Omega_1 \text{d}\Omega_2 \text{d}r_2 }, \label{eq:elec_dens}
\end{equation}
where the integral sums over the angular dependence as well as the radial part of one of the electrons. In the
large-$r$ region, $\rho(r_1,t)$ represents the electron density of the autoionizing electron. In this section, energy, time, and
distance are in electron Volts (eV), femtoseconds (fs), and Bohr radius (a.u.) respectively unless otherwise specified. The bandwidth
in energy and the duration in time are defined in FWHM.

\begin{figure}[htb]
\centering
\includegraphics[width=\linewidth]{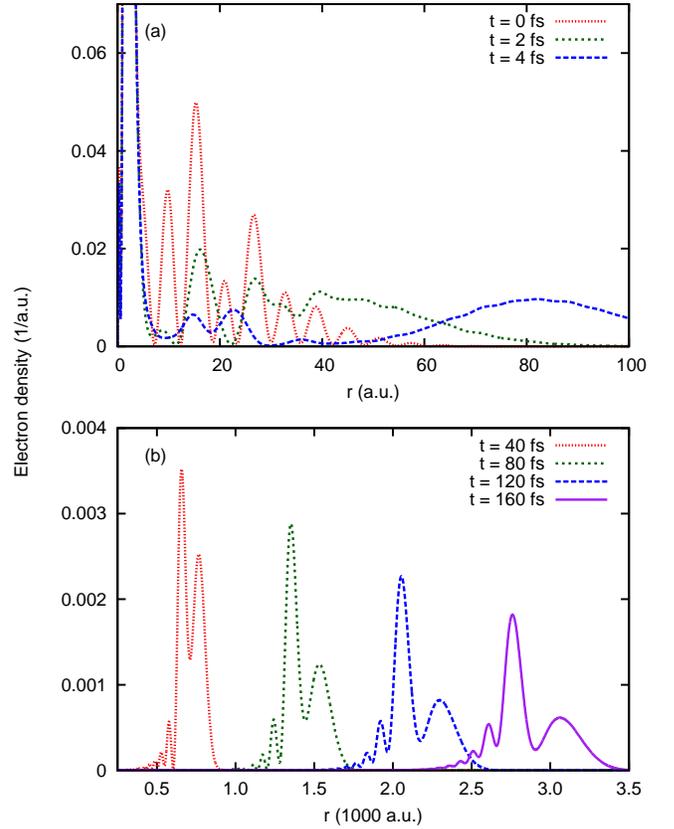}
\caption{The electron density for the $2p4s$ resonance in Be by a pump pulse of 2~fs and central frequency at the resonance. The
decay lifetime of the resonance is 3.78~fs. The electron density is shown for (a) short-time behavior at $t=0$, 2~fs, and 4~fs, and
for (b) long-time behavior at $t=40$~fs, 80~fs, 120~fs, and 160~fs.}
\label{fig:elec_dens}
\end{figure}

First we show how the wave packet moves in the coordinate space. Fig.~\ref{fig:elec_dens}(a) displays the electron density at $t=0$,
2~fs, and 4~fs, where the time scale is about the lifetime of the resonance $T=3.78$~fs. Before the propagation starts ($t=0$), a
dominant spatial distribution covers 0-60~a.u. This distribution is mainly contributed by the continuum, where the percentage of
initial bound state is as low as 7\%. The spatial oscillation with period of about 10-15~a.u., or wave number of about 0.4-0.6~a.u.,
matches the energy distribution centered at the resonance energy of 2.789~eV. As time goes on, the electron wave packet moves outward
toward large $r$. When time is long compared with the autoionization lifetime, the motion of the electron wave packet can be analyzed
more easily. In Fig.~\ref{fig:elec_dens}(b), the long-time behavior of the wave packet are plotted for $t=40$~fs, 80~fs, 120~fs,
and 160~fs. The electron is always confined in some spatial range, but it gradually spreads wider when time increases. By recognizing
and tracing the dominant two peaks in Fig.~\ref{fig:elec_dens}(b) over time, we find that the corresponding velocities are almost
constant. It is similar to a free electron wave packet traveling in space, where its average velocity keeps the same, but its spatial
distribution spreads. In this long time regime, the electron moves in a very predictable way to the large distance toward the
detector.

\begin{figure}[htb]
\centering
\includegraphics[width=\linewidth]{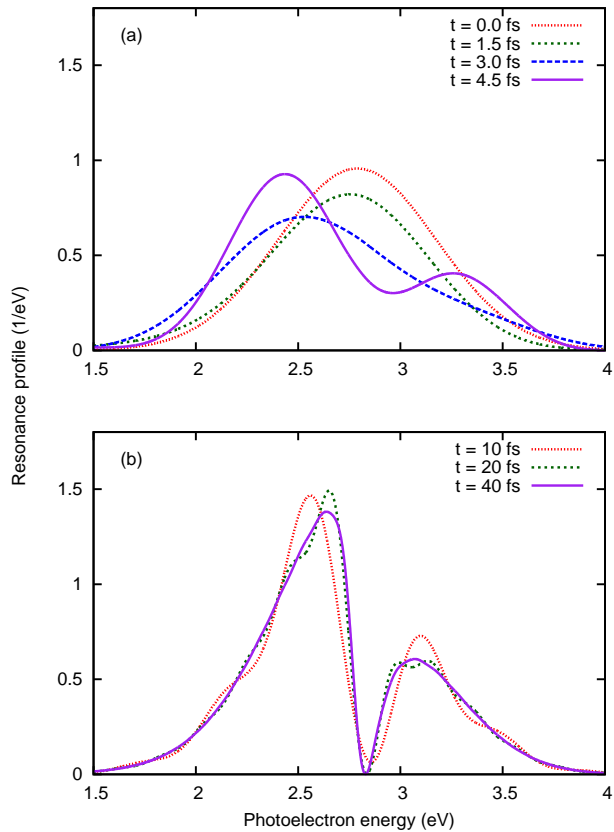}
\caption{The resonance profile $|d_{E'}(t)|^2$ for the $2p4s$ autoionization in Be, at (a) $t=0$, 1.5~fs, 3~fs, and 4.5~fs, and at
(b) $t=10$~fs, 20~fs, and 40~fs.}
\label{fig:profile}
\end{figure}

Next, we discuss the evolution of the resonance profile $|d_E(t)|^2$, i.e., the time-dependent Fano profile, for the same wave
packet in Eq.~\ref{eq:Psi_2p4s}. In Fig.~\ref{fig:profile}(a), the photoelectron energy profile is shown at $t=0$, 1.5~fs, 3~fs, and
4.5~fs, i.e., near the lifetime of the resonance $T=3.78$~fs. The profile starts evolving with a Gaussian shape at $t=0$ determined by
the pump pulse, and changes quickly within its lifetime. At $t=4.5$, just a little longer than the lifetime, the double-peaked
shape characteristic of autoionization can be seen, but the shape is not quite "settled" into the Fano profile. The long-time
behavior is shown in Fig.~\ref{fig:profile}(b). At $t=10$~fs to 20~fs, the resonance profile is already quite close to the final Fano
profile that can be determined in the standard energy domain measurements. It is also physically sensible that the electron velocities
observed in Fig.~\ref{fig:elec_dens}(b) can be mapped onto, although not exactly, the energies of the two highest peaks in
Fig.~\ref{fig:profile}(b).

\begin{figure}[htb]
\centering
\includegraphics[width=\linewidth]{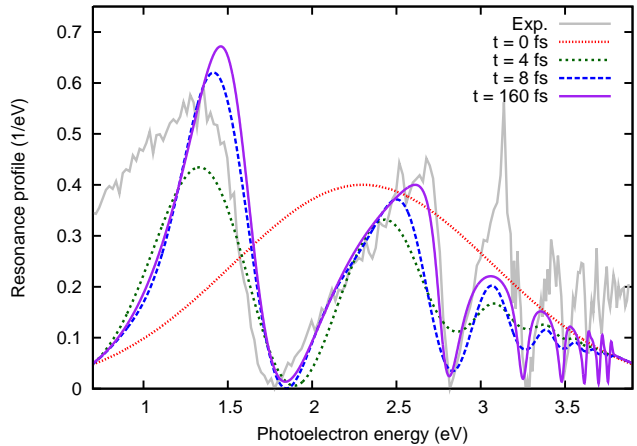}
\caption{The calculated resonance profiles for the $2pns \left( ^1 P ^o \right)$ resonances at $t=0$, 4~fs, 8~fs, and 160~fs
in Be, and the experimental photoabsorption cross section in Be between the $2s$ and the $2p$ thresholds~\cite{wehlitz}. The
 experimental values are normalized to the vertical range of the calculation.}
\label{fig:manyres}
\end{figure}

We have discussed an isolated resonance. Practically, a short pump pulse is likely to populate more resonances at once. Here we
consider the case including the $2pns$ resonances up to $2p9s$ in Be. Even though the $2p2s$ state (about 5.277~eV above the ground
state~\cite{nist}) is lower than the $2s$ threshold and will not show up in the continuum profile, it can in general interact with
the continuum beyond the $2s$ threshold and is included in the present calculation. The parameters for these resonances are generated by
setting $\mu = 0.6$, $\Gamma \times \nu^3=0.71$~eV, and $q=-0.8$, where $\mu$ is the quantum defect and $\nu$ is the effective
quantum number. The first few widths, starting from $2p3s$, are 0.514, 0.181, 0.0833, and 0.0451, in ~eV's, and the decay lifetimes
are 1.28, 3.64, 7.90, and 14.6 in femtoseconds, respectively. The $2pns$ resonance series stretches between the $2s$ threshold and
the $2p$ threshold, a range of about 4~eV. Our attention is directed at the time scale and energy scale of this series. The initial
continuum distribution is set to be centered at 2.3~eV above the ionization threshold, with bandwidth of 1.825~eV, or pump duration
of 1~fs. The bandwidth, covering the resonances of interest, is limited to the energy range above the $2s$ threshold to avoid
unnecessary complications. Each resonance term in
Eq.~\ref{eq:multi_cont} is multiplied by a Gaussian window whose width is 2$\Gamma$ to narrow down the interaction $V$ to some finite
range in the calculation. The resultant resonance profiles are displayed in Fig.~\ref{fig:manyres}, along with the experimental
photoabsorption cross section values in Ref.~\cite{wehlitz}, plotted by the light gray curve and normalized to fit in the vertical
range of our calculation. The $2pns$ resonance series is seen in the experiment from low to high energy, starting with $2p3s$ at
about 1.5~eV. For the calculation, the figure shows how the profiles evolve in time and are finalized, where the lower (wider)
resonances build up quickly and the higher (narrower) resonances slowly. At short times, the continuum must be viewed as a whole
distribution which gives no information on individual resonances; only at large times, e.g. $t=160$~fs, the profiles assigned to
individual resonances can be seen. Nonetheless, our calculation and the experiment in Ref.~\cite{wehlitz} should not be compared with
each other directly. In the present work, a short pump pulse covers a wide energy range; the system propagates coherently with
time. On the other hand, the experiment in Ref.~\cite{wehlitz} was done with many independent incoherent measurements, where each
measurement was made for a single energy point using a long pulse of at least of hundreds of picoseconds, and gives back a single
quantity for the cross section.

We caution the readers that the time-dependent resonance profile shown in Fig.~\ref{fig:manyres} is from theoretical calculations only,
i.e., a plot of $|d_E(t)|^2$ vs $E$ for different time delays after the pump pulse. As emphasized earlier, $|d_E(t)|^2$ cannot be measured
directly in the laboratory since any physical measurement will take at least many picoseconds which cannot distinguish time evolution
that happens on the femtosecond scale. Secondly, the wave functions associated with the coefficients $d_E(t)$ in Eq.~\ref{eq:Psi_multi}
are not eigenstates of the atom such that $|d_E(t)|^2$ cannot be directly interpreted as measurable probabilities.

% PROBE %%%%%%%%%%%%%%%%%%%%%%%%%%%%%%%%%%%%%%%%%%%%%%%%%%%
\section{Method of probing the time-dependent autoionization processes}
\label{sec:probe}

After we have predicted how the wave packet in the autoionization process evolves with time, an experimental scheme where such
predictions can be confirmed is essential. Using SAP-pump and IR-probe, the lifetime of Auger decay of an inner-shell
hole~\cite{drescher} or of a Fano resonance in helium~\cite{gilbertson} have been "deduced". The streaking of an autoionizing
electron in the IR laser field has been calculated by Wickenhauser \textit{et al}~\cite{wickenhauser}. The streaked electron spectra
are too complicated to allow the retrieval of the predicted intermediate profiles. Taking beryllium as example, here we propose in
the following a way to measure the time dependence of Fano profiles.

\begin{figure}[htb]
\centering
\includegraphics[width=\linewidth]{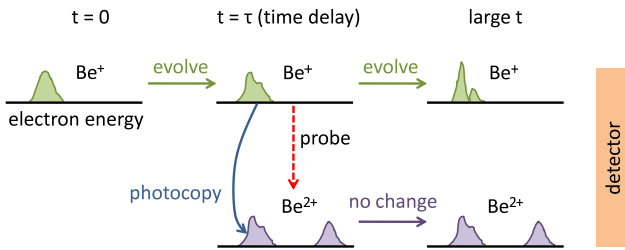}
\caption{A conceptual scheme of probing the time-dependent resonance profiles. At $t=0$ the pump pulse is over and autoionization
starts while the system evolves freely. To find out the the evolving electron wave packet at time  $\tau$, a probe pulse is used to
ionize the $2s$ electron in the $2sEp$ continuum and create the $E'pEp$ states. After the probe is over, the new wave packet consists
of a part of the old wave packet which continues to evolve and autoionize (the upper row) and the newly generated part made of $E'pEp$
states (the lower row). The latter does not change with time. By picking the responsible energy range and collecting the double
ionized signals, the resonance profile in the autoionization at the moment $\tau$ is captured.}
\label{fig:probescheme}
\end{figure}

The probe scheme is sketched in Fig.~\ref{fig:probescheme}. After the pump, we define the starting time for the autoionization
as $t=0$ such that the pump field is finished for $t>0$. A probe pulse is shined onto the system with its peak at time delay $\tau$.
The probe has the duration of 1.5~fs, or bandwidth of 1.216~eV, and carrier energy $\omega$ of 40~eV. This probe pulse ionizes the
$2s$ electron in the $2sEp$ continuum (binding energy of $2s$ is 18.21~eV). After the probe pulse is over, the excited wave packet of
Eq.~\ref{eq:Psi_2p4s} changes to
\begin{align}
\left| \Psi_{ex}(t) \right\rangle = \sum_n { d_n(t) | 2pns \rangle } + \int{ \bar{d}_E(t) | 2sEp \rangle \text{d}E } \notag\\
+ \iint{ d_{E'E}(t) | E'pEp \rangle \text{d}E'\text{d}E }.
\end{align}
We have assumed that the $2pns$ bound states are not changed by the probe pulse, and only a small part of the $2sEp$ continuum states
are ionized (thus the amplitude now given by $\bar{d}_E(t)$ is slightly different from $d_E(t)$) to generate new $E'pEp$ states. After
the probe pulse, both $d_n(t)$ and $\bar{d}_E(t)$ continue to change with time as the bound states decay. The newly created
$E'pEp$ states are eigenstates of the atomic (field-free) Hamiltonian and are stationary states, and they can thus
be probed by a laboratory detector. Using 40 eV probe pulses, the energy $E'$ will be centered around 21.8~eV, well-separated from the
electron energy of 2.5~eV due to autoionization. Clearly the probability $|d_{E'E}(t)|^2$ should be independent of time and is
proportional to $|d_E(\tau)|^2$. To determine $|d_E(\tau)|^2$ one can measure the
energies of the two electrons in coincidence or by measuring the low-energy electrons in coincidence with Be$^{2+}$ ions. Either way
lower energy group duplicates the original continuum profile at the moment of the probe. By changing the delay time of the probe a
copy of $|d_E(\tau)|^2$ can be determined. Due to the finite duration of the probe pulse, the measured copy of
$|d_E(\tau)|^2$ will be somewhat smoothed out.

\begin{figure}[htb]
\centering
\includegraphics[width=\linewidth]{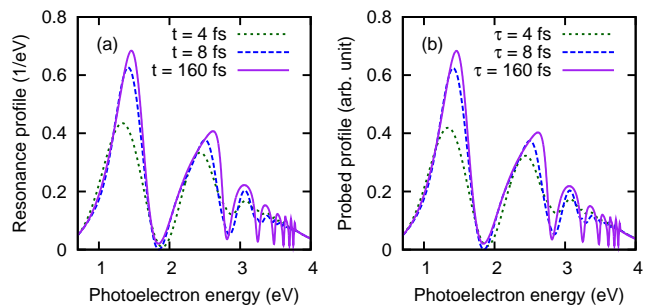}
\caption{The time-dependent continuum profile compared with the probed profile in the Be case. (a) The calculation as specified in
Fig.~\ref{fig:manyres}. (b) The retrieved profile using a probe beam with $\omega=40$~eV and the duration (FWHM) of 1.5~fs. The
signal values have been scaled to match the original profile at $t=160$~fs. The agreement between (a) and (b) is good.}
\label{fig:probe}
\end{figure}

If the probe is of perturbative strength, we can calculate the predicted two-electron spectra quantitatively. In the presence of
the external probe field $\vec{F}_2(t)$, the coefficient for $E'pEp$ is given by the first-order time-dependent perturbation theory by
\begin{align}
d_{E'E} &= \left\langle E'pEp \left| U \left( \frac{T_0}{2}, -\frac{T_0}{2} \right) \right| \Psi_{ex} \left( -\frac{T_0}{2} \right)
\right\rangle \label{eq:probe1}\\
&= \frac{1}{i} \int_{-\frac{T_0}{2}}^{\frac{T_0}{2}}{ \left\langle E'pEp \left| e^{-i H_0 \left( \frac{T_0}{2} - t \right)}
V_2(t) \right| \Psi_{ex}(t) \right\rangle \text{d}t } \label{eq:probe2}\\
&= \frac{\mu}{i} \int_{-\frac{T_0}{2}}^{\frac{T_0}{2}}{ e^{ -i (E'+E) \left( \frac{T_0}{2} - t \right)} F_2(t) d_{E'} (\tau + t)
\text{d}t } \label{eq:probe3}
\end{align}
where $V_2(t)$ is the dipole interaction potential for the probe pulse. The interaction is turned on only for
$-T_0/2 < t < T_0/2$. Note that $t$ is now measured with respect to the center of the probe, and the total excited wave function
$| \Psi_{ex} (-T_0/2) \rangle$ at $t=-T_0/2$ has already gone through the autoionization process for a duration of $\tau - T_0/2$. In
Eq.~\ref{eq:probe1}-\ref{eq:probe2}, the time-propagator $\exp{ [-iH_0 (T_0/2+t) ] }$ in $U$ operates on $\Psi_{ex} (-T_0/2)$ and
gives $\Psi_{ex} (t)$ in return. In the next step in Eq.~\ref{eq:probe3}, since the bound states $2pns$ do not participate in the probe
process, only the continuum part is written down. The dipole matrix element $\mu \equiv \langle E'p |z| 2s \rangle$ is assumed to be
a constant. The polarization of the probe field is in the $z$-direction, i.e. $\vec{F}_2(t) = \hat{z} F_2(t)$. The retrieved profile as
a function of $E$ is defined as $\int{ | d_{E'E} |^2 \text{d} E' }$. Fig.~\ref{fig:probe} shows very good agreement between the
retrieved profile at different $\tau$'s and the original continuum profile at the associated time $\tau$.

While the calculation given above is in first-order perturbation, the scheme can be applied to intense probe as well as long as the
probe field dominantly removes the $2s$ electron and has little effect on other components in the system. The $Ep$ electron is not
affected by the probe, and the continuum profile of $2sEp$ at $\tau$ will still be transferred to $E'pEp$ plainly (with additional
peaks possibly from two-photon ionization). The advantage of a stronger probe is that it increases the signal rate. This may be an
important factor to realize this pump-probe scheme since the population of double excited states by the pump can be already weak. In
the present example, the photon energy is 11.62~eV for the pump and 40~eV for the probe, presuming that the short UV sources
are provided for both the pump and the probe.

% CONCLUSIONS %%%%%%%%%%%%%%%%%%%%%%%%%%%%%%%%%%%%%%%%%%%%%%%%%%%
\section{Conclusions}
\label{sec:conclusions}

In this paper we study the temporal behavior of the atomic autoionization process. Following the original paper of Fano, we treat
autoionization as the decay of a discrete state into a degenerate continuum. Fano's theory characterizes such a resonance with a
shape parameter $q$ and a width $\Gamma$, in addition to the resonance energy position. For many decades, Fano resonances are
measured and characterized in energy domain experiments. With the advent of XUV and soft X-rays with pulse durations of attoseconds
to a few femtoseconds, we address the question whether Fano resonances can be probed directly in the time domain within its typical
lifetime. In particular, can we talk about the time-dependent Fano profiles before the discrete state is fully decayed and can such
profiles be measured experimentally?

To describe time-dependent Fano profiles, it is inconvenient to express the total time-dependent wave function in terms of the
eigenstates of the system. Instead, the configuration basis states, including the bound and the continuum, used in Fano's original
theory are much more useful (or more physical) for analyzing the autoionization process, and we are led to define the time-dependent
Fano profiles by referring to the wave packets expanded in the continuum configuration basis states. To probe the temporal Fano
profiles with good time resolution, we suggest that the decay process be perturbed by a short soft-X-ray pulse which interrupts
the autoionization at the time of the probe. The probe in the meantime creates a stationary wave packet that can be measured in the
laboratory. We further consider many Fano resonances together if they fall within the bandwidth of the pump pulse.

The pump-probe scheme proposed here, requiring high-energy short pulses for both the pump and the probe, is not yet available today.
On the other hand, attosecond technology is undergoing rapid growth, the proposed scheme of measuring autoionization dynamics may
become doable within a few years. At the same time,  theory for attosecond electron dynamics is still a barren field today. General
approaches for understanding electron dynamics for many-electron systems are badly needed at this time. Not all the probe pulses have
similar effectiveness in retrieving the time-dependent wave packets. Without the general theoretical framework, it would be difficult
to retrieve electron dynamics from the signals measured by the probe pulse.

\begin{acknowledgments}
This work was supported in part by Chemical Sciences, Geosciences and Biosciences Division, Office of Basic Energy Sciences, Office
of Science, U.S. Department of Energy.
\end{acknowledgments}

\end{document}